\documentclass[12pt]{article}
\usepackage{pdproc, epsfig}

\begin{document}

\renewcommand{\thefootnote}{\alph{footnote}}

\title{
 A UNIQUE DETECTOR FOR PROTON DECAY AND NEUTRINO OSCILLATIONS
 STUDY (LANNDD) FOR A USA DUSEL\footnote{Invited talk at the XI Neutrino Telescope
 Conference,
 February 2005, Venice, Italy; parts of this paper are
 from a white paper submitted to SAGENAP, http://www.physics.ucla.edu/hep/whitepaper/wp.pdf.}}

\author{David B. Cline}

\address{Astrophysics Division
Department of Physics \& Astronomy,\\ University of California, Los Angeles\\
 Los Angeles, CA 90095, USA\\
 {\rm E-mail: dcline@physics.ucla.edu}}

  \centerline{\footnotesize and}

\abstract{We discuss the major scientific issues of the search for
proton decay to $10^{35}$ years lifetime and search for CP
violation with a VLBL superbeam (~2000 km distance). The 100 kT
LANNDD liquid Argon TPC is well matched to these goals. We
describe the progress in the R\&D program for the detector as well
as the possible location in an underground laboratory in the USA
called DUSEL.}

\normalsize\baselineskip=15pt

\section{Outline}
1.  The DUSEL concept

2.  The development of the ICARUS liquid Argon detector

3.  The LANNDD detector concept and current worldwide R\&D effort

4. Sensitivity to proton decay and VLBL neutrino beam CP violation
with $\nu _{\tau} \rightarrow \nu _{e} $

5.  Safety issues at DUSEL for LANNDD

6.  Summary

\section{Introduction}

The major scientific issues of elementary particle physics may be
partially solved with the CMS and Atlas detector operation at the
LHC. Other problems such as the nature of dark energy and dark
matter are under investigation around the world (see the
proceedings of The 6th Symposium on the Sources and Detection of
Dark Matter and Dark Energy in the Universe, Marina Del Rey,
California, 2005, in press).

Other equally fundamental questions involve the possibility of
proton decay and CP violation in the neutrino sector. Until now
these areas of science have been dominated by the use of water
Cherenkov detectors such as Super K and IMB. However, with the
successful development of the ICARUS liquid Argon TPC for the Gran
Sasso Laboratory a new very powerful detector technique has
entered the field.

In order to mount a new method to search for proton instability
(mainly $p \rightarrow k ^+ \overline{\nu}$ a new very large
underground laboratory is needed. It is possible that such a
laboratory will be developed in the USA called DUSEL (Deep
Underground Science and Engineering Laboratory) by the U.S.
National Science Foundation. In this report we discuss the
progress to develop a 100 kT liquid Argon detector called LANNDD
[1][2]. This detector is well matched to the size and scope of
DUSEL and a possible super neutrino beam produced either by BNL or
FNAL. A powerful R\&D program for LANNDD is in progress that we
will describe.

\section{The DUSEL Concept}

In the USA there is a plan to develop a deep laboratory for large
modules (proton decay detector) and small modules (double beta
detectors). Eight proposals have been to NSF. UCLA is the
principal investigator for Carlsbad proposal. There are five
co-PIs including the author. A key concept is the use of a
neutrino beam from BNL or FNAL VLB to one of these sites (to be
discussed later). The sites are shown in Figure 1 [3].

\section{The Status if the ICARUS Liquid Argon Detector}

The ICARUS LAR detector has made great progress recently [4].

(a) The T600 has now moved to the LNGS Hall B

(b) Work is starting to construct the next T1800 module

(c) The CNGS neutrino beam will be sent to the LNGS mid-2006;
ICARUS will detect.

With about 2 kT of detector and the R \& D program discussed here
the time is ripe to plan a larger $\approx $ 100 kT detector.
LANNDD is a major candidate for this detector and it could be
located in DUSEL.

\section{The LANNDD Concept: Possible application of this technology in the
USA}

One option for a next generation nucleon decay search instrument
is a fine-grained detector, which can resolve kaons as well as
background from cosmic ray neutrinos that are below the threshold
for water Cerenkov detectors such as Super-Kamiokande (SK). Such a
detector can make progress beyond the few $ \times 10^{33}$ year
limits from SK for SUSY favored modes because the reach improves
linearly with the time and not as the square root of exposure as
in SK. It will be possible to discover nucleon decay up to about
$\approx 10^{35}$ year lifetime/branching ratio with an instrument
of $\approx $ 70 kT mass in liquid argon, as the one studied in
the LANNDD project after a few years of exposure [5].

A second major goal for such an instrument, as demonstrated in a
spectacular example of synergy in the last two generations of
underground detectors, is the study of neutrino interactions and
oscillations [6]. Such a detector can make neutrino oscillation
studies using the cosmic ray neutrinos alone (being able to
resolve muon neutrino regeneration, detect tau's and tighten
measurements of $ \Delta m^2 $ search for other mixing than $\nu
_{\tau} \rightarrow \nu _{e}$. But coupled with a neutrino
factory, this detector, outfitted with a large magnet, offers the
advantage of being able to discriminate the sign not only of muon
events, but of electron events as well. Given the
bubble-chamber-like ability to resolve reaction product
trajectories, including energy/momentum measurement and excellent
particle identification up to a few GeV, this instrument will
permit the study of the neutrino MNS matrix in a manner that is
without peer. The LANNDD detector is shown in Figure 3.

One may question whether such a marvelous instrument is
affordable, by which we mean buildable at a cost comparable or
less than the neutrino source cost. It is indicated, by simple
scaling from existing experience with ICARUS, that such an
instrument will cost out in the class of a large collider detector
instrument and represents a straightforward extrapolation of
existing technology.

As expected for such a large, isotropically sensitive,
general-purpose detector, there are many ancillary physics goals
that can be pursued. This device would allow exploration of
subjects ranging from the temporal variation of the solar neutrino
flux (above a threshold of perhaps 10 MeV), to searches for
neutrinos from individual or the sum of all supernovae and other
cataclysmic events (e.g. GRBs), to cosmic ray research
(composition, where the WIPP depth is advantageous), dark matter
searches (via annihilation neutrinos), searches for cosmic exotic
particles (quark nuggets, glueballs, monopoles, free quarks), and
point source neutrino astronomy. In all these instances, we can go
beyond SK by virtue of lower energy threshold, better energy loss
rate resolution, momentum, angle, sign and event topology
resolution.

\section{Sensitivity to Proton Decay etc.}

Much of the scientific studies that are being done with LANNDD
follow the success of the ICARUS detector program. The main
exception is for the use of the detector at a neutrino factory
where it will be essential to measure the energy and charge of the
m± products of the neutrino interaction [3]. We will soon propose
an R\&D program to study the effects of the magnetic field
possibilities for LANNDD.

\subsection{Search for Proton Decay to $10^{35}$ Years}

The detection of $p \rightarrow k ^+ \overline{\nu} _{\mu} $ would
seem to be the key channel for any SUSY model. This channel is
very clear in liquid argon due to the measurement of the range and
detection of the decay products. We expect very small background
events at $10^{35}$ nucleon years for this mode (refer to ICARUS
studies) [6].

\subsection{Solar Neutrinos and Supernova Neutrinos Studies}

The major solar neutrino process detected in liquid argon is $\nu
_{e} + {}^{40} Ar \rightarrow {}^{40}Ar ^* + e^-$, with $Ar^{*}$
de-excitation giving photons with subsequent Compton events. The
same process is useful for supernova $\nu _{e} $ detection - the
expected rate for the solar neutrinos is  $\approx $ 123,000 per
year. For a supernova in the center of the galaxy with full mixing
there would be ~3000 events - no other detection would have this
many clear $\nu _{e} $ events.

\subsection{Atmospheric Neutrino Studies}

By the time LANNDD is constructed it is not clear which
atmospheric neutrino process will remain to be studied. However
this detector will have excellent muon, hadron and electron
identification as well as the sign of $\mu \pm $ charge. This
would be unique in atmospheric neutrino studies. The rate of
atmospheric neutrinos in LANNDD will be (50 kT fiducial volume):

CC $\nu _{e} $ events: 4800/year

CC $\nu _{\mu } $ events: 3900÷2800/year (depending on the
neutrino mixing).

There would also be about 5000 NC $\nu$ events/year. We would
expect about 25 detected $\nu _{\tau } $ events/year that all
would go upward in the detector.

\subsection{Use of LANNDD in a VLBL Beam}

Because of the large mass and nearly isotropic event response,
LANNDD could observe neutrinos from any of the possible neutrino
beam: BNL, FNAL. The LANNDD detector could be useful for the
search for CP violation from any BNL or FNAL beam. This will
depend on the value of the mixing angle $\theta _{13} $ and the
magnitude of the CP violation. Figure 1 shows the distance to all
the DUSEL sites.

\section{Safety Issues for LANNDD at DUSEL}

The safe installation and operation of the proposed Large Liquid
Argon Detector, with tens of thousands tons of oxygen replacing
liquid argon at cryogenic temperatures, will require significant
safety engineering considerations.  Siting a facility such as the
LANNDD in a hard rock location, where mining is difficult and
slow, could add significantly to the overall cost of supplying
suitable egress and suitable venting systems in the event of major
disruptions of containment.  By contrast, the low cost of mining,
shaft installation, and material removal in a salt mine provides a
cost-effective solution to the safety implications posed by
installation and operation of LANNDD in an underground
environment.  Multiple emergency air dumps and sources, along with
multiple egress options, can be simply created at reasonable cost.

The safety installations described above can be obviously applied
also to other underground locations, with some specific
adaptations and modifications of the solutions to the local and
environmental conditions.

The WIPP facility has a long history of safe operation;  it can
claim to be the safest underground operation in the U.S.  today
and has already won numerous safety awards.

The safety study will include the following items (for a more
detailed description of these items see the SAGENAP report web
site on the first page. Engineering Feasibility and Safety Study
for LANNDD and Other Particle Detectors at WIPP):

\section{R\&D effort: 5m Drift Experiment (this section prepared with Franco Sergiampietri, Pisa}

\subsection{Introduction}

Aimed to the fields of neutrino physics and of nucleon decay, a
preliminary study was started, in the year 2000, for identifying
the main configuration criteria, the critical items and the
related required tests for a 50-100 kTon detector, configured as
liquid argon TPC, eventually fitted out with magnetic field. This
study pointed out general thoughts, dictated by detection
efficiency, mechanical stability, feasibility, underground
operation, safety and costs, for the choice of:

- shape

- modularity

- detector parameters

- cryostat parameters

- magnet parameters

- site parameters.

The results, presented at NuFact'01, Tsukuba , describe a possible
future detector named LANNDD (Liquid Argon Neutrino and Nucleon
Decay Detector), sited at the WIPP site, Carlsbad, NM. A
huge-scale detector, as the conceived one, requires a preliminary
R\&D activity for directly verifying all a series of design
parameters, peculiar of its large scale, beyond the indications
obtainable from extrapolation from much smaller size detectors or
from Monte Carlo simulations. In particular:

- long electron drift (4-8 m)

- track reconstruction in magnetic field (e.m. shower sign
discrimination, muon momentum resolution),

- electron drift velocity at different (high) hydrostatic
pressures should be considered as first priority tests.

In the present article we propose an experimental set-up for the
first of these tests.

\subsection{Long Electron Drift in Liquid Argon}

In order to decrease the complexity and the cost of a large scale
detector it is important minimizing the number of electron
collection wire planes. The purpose of the test is to
experimentally verify the limits set on the maximum drift length
by diffusion and attenuation of the drifting charge, by measuring
the electron collection efficiency with a 5m (or greater) drift
path. For a long drift space the problems to be faced are:

-   Use of high voltages for the drift in the range $V _d \approx
200 - 400 kV $. For a drift field $E_d = 0.5kV/cm $ and a drift
space $d = 5m $, $V_d = 250 kV $ is required. The maximum drift
time is $T_d = 3.1 ms $ . Every detail of the high voltage system
should be carefully designed for a safe operation in a low noise
environment as the liquid argon TPC.

\subsection{Experimental Methods}

We plan to measure attenuation and diffusion effects for ionized
tracks along a 5-m long drift space. A first method is based on
recording tracks induced by vertical cosmic ray, selected at
different distance along the drift by a pair of scintillator
counters in coincidence. After each trigger from the counters,
data are acquired during a time window corresponding to the
maximum drift time. The dependence on the counter position of the
peak of the pulse height distribution in the collection wires will
show the attenuation along the drift. The time duration of signals
is related to the diffusion along the drift direction.

A second method is based on recording single long tracks, slightly
inclined with respect to the drift direction, from a muon test
beam (pµ ³ 10 GeV/c). Portions of the track originated at
different distance from the collection wire plane are collected in
different wires and their pulse height distribution gives
immediately the exponential behavior due to the attenuation.

\subsection{The Detector}

The detector is configured as a time projection chamber with
cylindrical drift volume. The drift volume, 5m long, is bounded by
a cathode, at one end, and by a wire chamber, at the opposite end.
The electric field generated between cathode and wire chamber, is
kept uniform by mean of a stack of uniformly spaced metallic rings
(field shaping electrodes), biased at voltages linearly decreasing
from the cathode voltage to the wire chamber voltage.

In a first configuration, the wire chamber is made by two wire
planes. The first plane, with grounded horizontal wires, works as
grid (Fisher type) for screening the second wire plane (collection
plane) from the drift volume and then avoiding position dependent
signals. The collection wires are vertically oriented and
individually connected by signal lines to a low voltage
multi-contact feedthrough and, from there, to the outer front-end
electronics (charge sensitive preamplifiers, amplifiers, ADC's).

The current collaboration is between Pisa, Granada and UCLA.

\section{Acknowledgements}
  I wish to thank Franco Sergiampietri for help and E. Fenyves for
the Safety Study and all members of the WIPP/DUSEL team for help
and discussion.

\begin{figure}
\vspace*{13pt} \leftline{\hfill\vbox{\hrule width 5cm
height0.001pt}\hfill} \mbox{\epsfig{figure=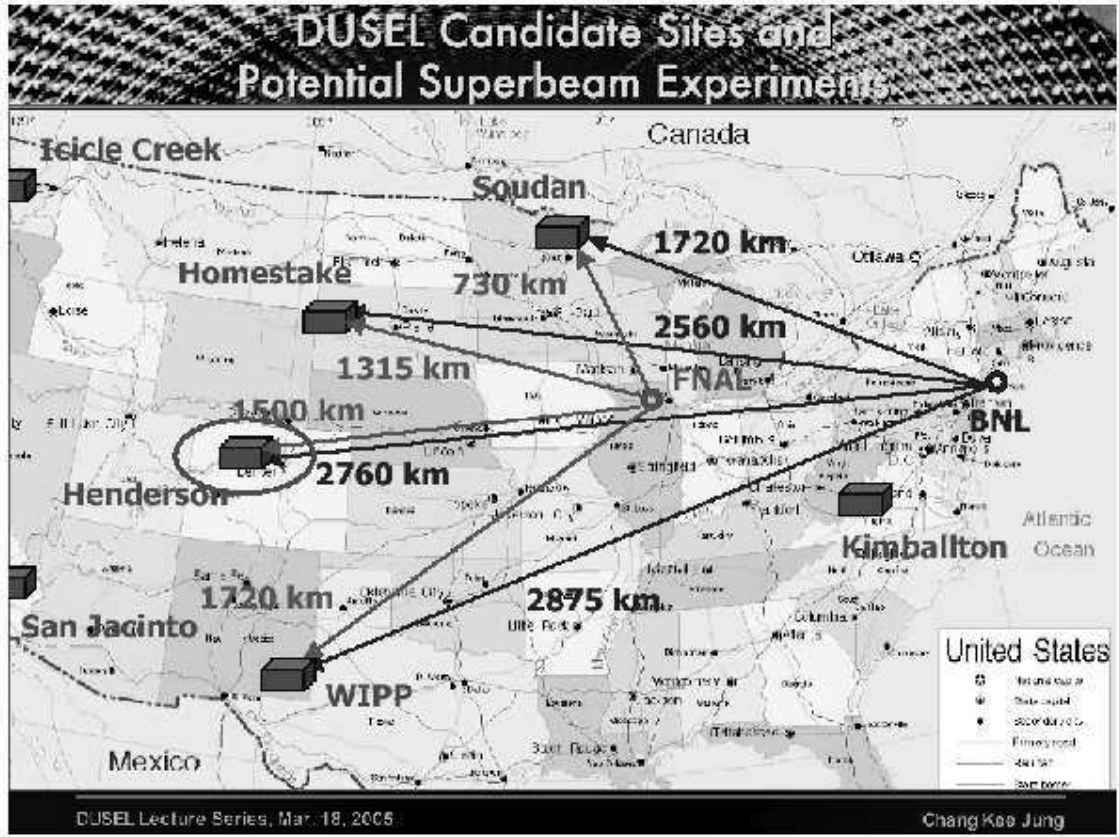,
width=15.0cm}}
\vspace*{1.4truein}     
\leftline{\hfill\vbox{\hrule width 5cm height0.001pt}\hfill}
\caption{DUSEL candidate sites.} \label{fig:fig1}
\end{figure}

\begin{figure}
\vspace*{13pt} \leftline{\hfill\vbox{\hrule width 5cm
height0.001pt}\hfill} \mbox{\epsfig{figure=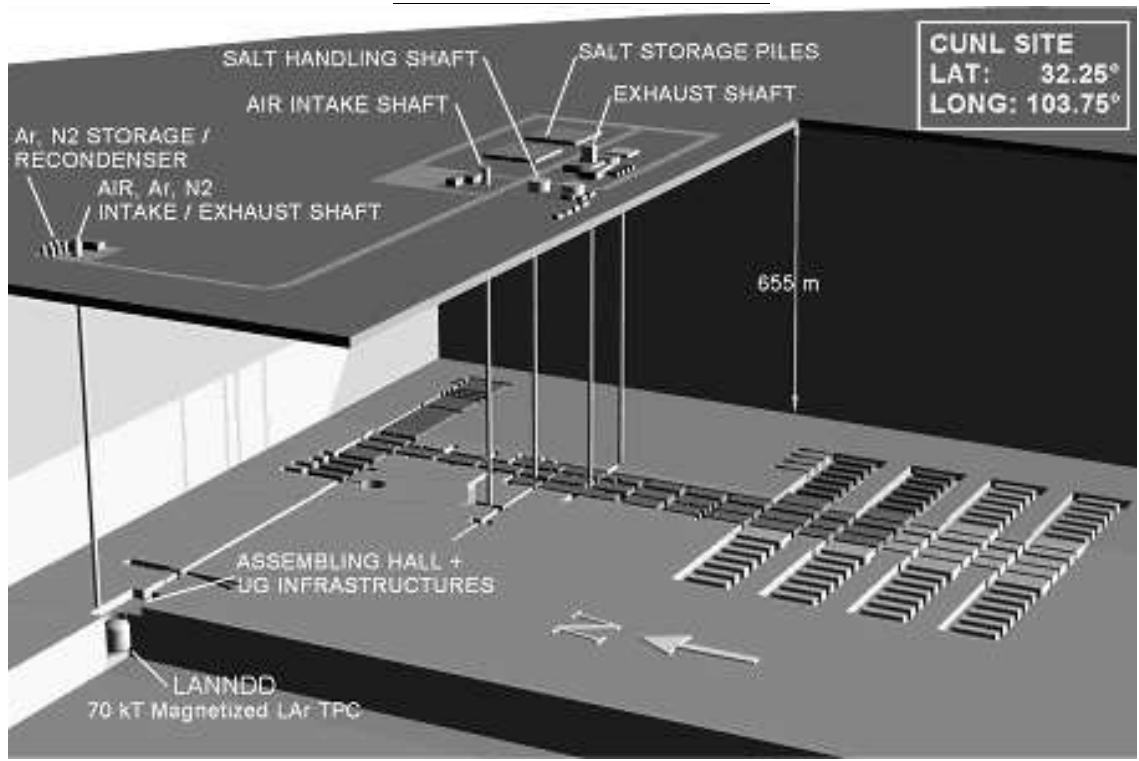,
width=15.0cm}}
\vspace*{1.4truein}     
\leftline{\hfill\vbox{\hrule width 5cm height0.001pt}\hfill}
\caption{LANNDD at the WIPP site.} \label{fig:fig1}
\end{figure}

\begin{figure}
\vspace*{13pt} \leftline{\hfill\vbox{\hrule width 5cm
height0.001pt}\hfill} \mbox{\epsfig{figure=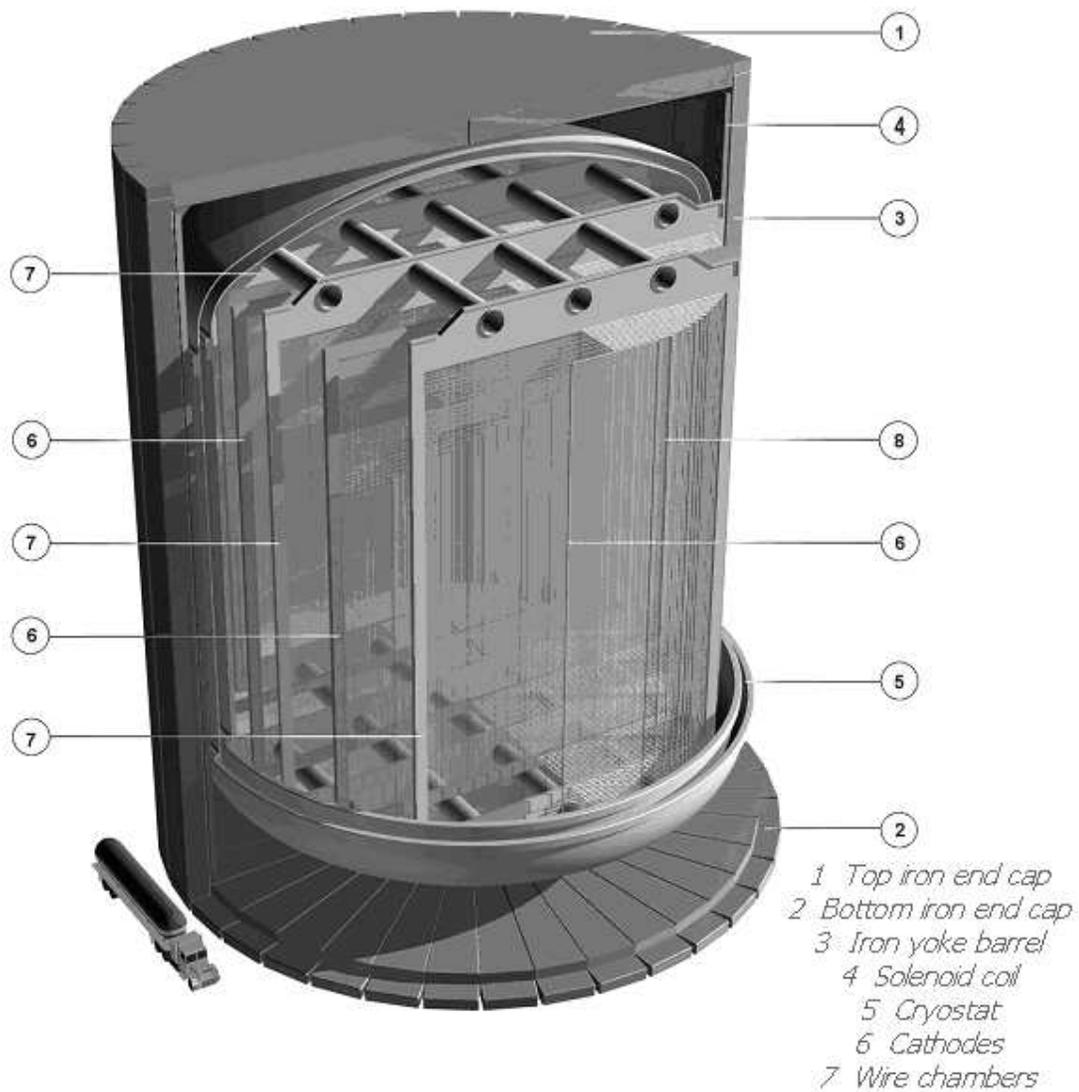,width=15.0cm}}
\vspace*{1.4truein}     
\leftline{\hfill\vbox{\hrule width 5cm height0.001pt}\hfill}
\caption{Cutaway view of the LANNDD detector.} \label{fig:fig3}
\end{figure}

\begin{figure}
\vspace*{13pt} \leftline{\hfill\vbox{\hrule width 5cm
height0.001pt}\hfill} \mbox{\epsfig{figure=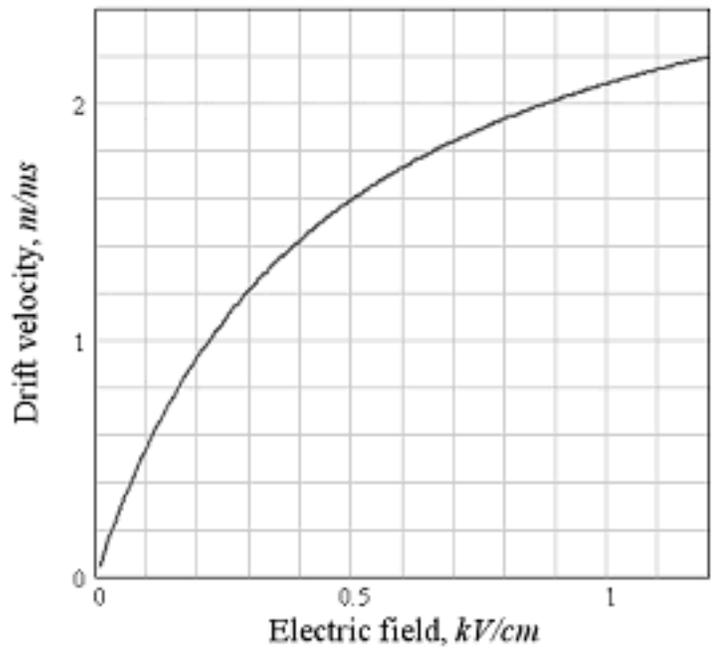,}}
\vspace*{1.4truein}     
\leftline{\hfill\vbox{\hrule width 5cm height0.001pt}\hfill}
\caption{Drift velocity versus electric field in liquid Argon.}
\label{fig:fig4}
\end{figure}

\begin{figure}
\vspace*{13pt} \leftline{\hfill\vbox{\hrule width 5cm
height0.001pt}\hfill} \mbox{\epsfig{figure=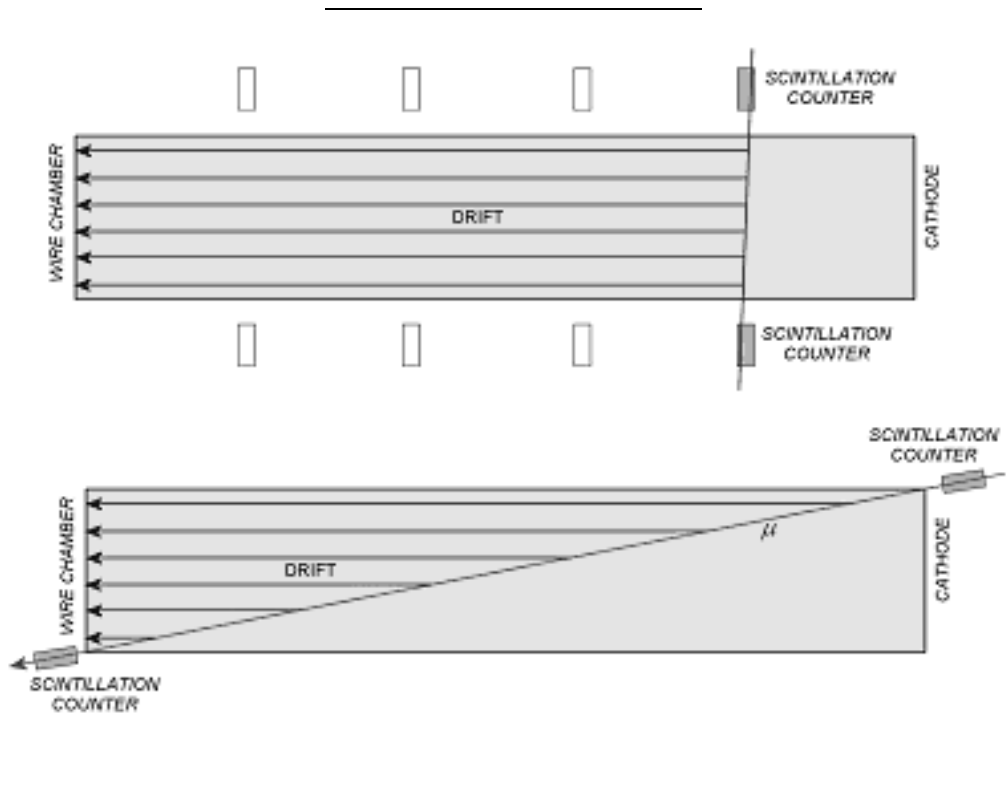,}}
\vspace*{1.4truein}     
\leftline{\hfill\vbox{\hrule width 5cm height0.001pt}\hfill}
\caption{Collecting event with vertical cosmic ray trigger (top)
and Collecting event with muon beam inclined with respect to the
drift direction (bottom).} \label{fig:fig5}
\end{figure}

\begin{figure}
\vspace*{13pt} \leftline{\hfill\vbox{\hrule width 5cm
height0.001pt}\hfill} \mbox{\epsfig{figure=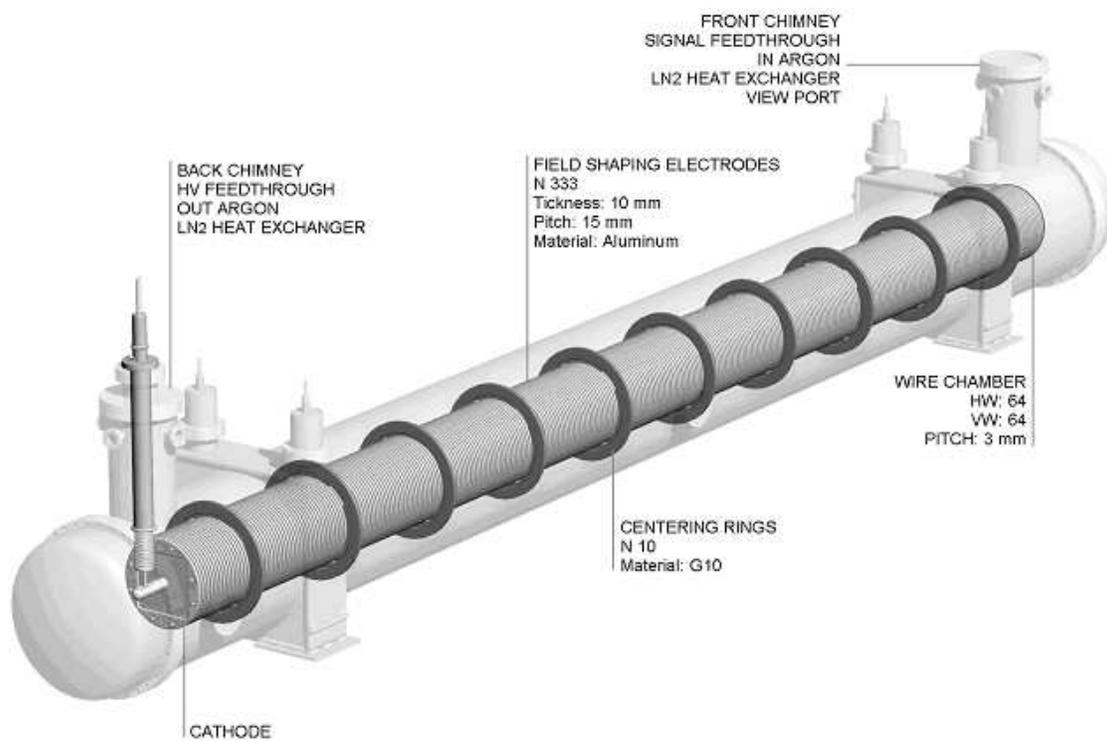,width=15.0cm}}
\vspace*{1.4truein}     
\leftline{\hfill\vbox{\hrule width 5cm height0.001pt}\hfill}
\caption{The 5-m drift time projection chamber.} \label{fig:fig6}
\end{figure}

\end{document}